%% file: manuscript.tex
\documentclass[twoside]{article}

\usepackage{graphicx}
\usepackage{bm}

\usepackage{cmpj2e}

\title{Colloidal particles in liquid crystal films and at interfaces}
\author{Mykola Tasinkevych\refaddr{label1,label2} and Denis Andrienko\refaddr{label3}}
\addresses{
\addr{label1} Max-Planck-Institut f\"{u}r Metallforschung, Heisenbergstr. 3, D-70569 Stuttgart, Germany\\email: miko@mf.mpg.de
\addr{label2} Institut f\"{u}r Theoretische und Angewandte Physik, Universit\"{a}t Stuttgart, Pfaffenwaldring 57, D-70569 Stuttgart, Germany
\addr{label3} Max-Planck-Institut f\"{u}r Polymerforschung, Ackermannweg 10, D-55128 Mainz, Germany\\email: denis.andrienko@mpip-mainz.mpg.de }

\newcommand{\unit}[1]{\ensuremath\mathrm{#1}}

\begin{document}

\maketitle

\begin{abstract}
\noindent This mini-review discusses the recent contribution of theoretical and computational physics as well as experimental efforts to the understanding of the behavior of colloidal particles in confined geometries and at liquid crystalline interfaces. Theoretical approaches used to study trapping, long- and short-range interactions, and assembly of solid particles and liquid inclusions are outlined. As an example, an interaction of a spherical colloidal particle with a nematic-isotropic interface and a pair interaction potential between two colloids at this interface are obtained by minimizing the Landau-de Gennes free energy functional using the finite-element method with adaptive meshes.
\pacs 61.30.Cz, 61.30.Jf, 61.20.Ja, 07.05.Tp
\end{abstract}

\input{introduction}
\input{theory}
\input{single_particle}

\input{pair_interactions}
\input{SmC_films}
\input{nematic_cells}
\input{interfaces}

\input{results}

\input{conclusions}

\section*{Acknowledgements}
D.A. acknowledges the Multiscale Materials Modelling Initiative of the Max Planck Society, DFG via IRTG program between Germany and Korea, DFG grants AN 680/1-1 and SPP1355, and BMBF grant MESOMERIE. M.T. acknowledges fruitful and stimulating discussions with Nuno Silvestre. Mara Jochum is acknowledged for critical reading of the manuscript.  We are grateful to Andrij Trokhymchuk for patience when collecting contributions for this issue.

\label{last@page}
\end{document}

%% file: introduction.tex
\section{Introduction}
Liquid crystals are known for their anisotropic mechanical and optical properties which originate from the long-range orientational molecular ordering. If a liquid crystal is used as a host liquid in a colloidal suspension, this ordering gives rise to additional long-range interactions between the colloidal particles. The type of the interaction is controlled by the presence and symmetry of topological defects of the director field. Particle clustering, formation of superstructures, and even new phases are immediate consequences of these anisotropic interactions.

In what follows, we present a brief review of theoretical and simulation methods which are used to study liquid crystalline systems. Each description of phenomena occurring on a particular length-scale has its own order parameter, such as the director field, tensorial order parameter, or particle density. For example, long-range forces between colloidal particles can be calculated by direct integration over the director field around them. For smaller particle-particle separations, where nonlinear effects become important and topological defects change their positions, a direct minimization of the Landau-de Gennes free energy functional with respect to the tensorial order parameter can be used. For even smaller separations, presmectic ordering and density variations cannot be ignored and density functional approaches or computer simulations can be employed.

With these techniques at hand, we discuss different physical situations: (i) a single particle suspended in a nematic host; (ii) an effective pair interaction of two particles suspended in a bulk nematic; (iii) an interaction of a particle with a nematic-isotropic (NI) interface and (iv) an interaction of two colloidal particles captured by this interface. We also give an overview of colloidal self- and directed-assembly in two-dimensional systems, such as free-standing smectic films and quasi two-dimensional nematic colloids. Finally, to give an example, we calculate the expelling force on a spherical particle in the vicinity of the NI interface and the effective interaction potential between two spherical particles captured by this interface. In both cases we assume the film-like shape of the domain occupied by the nematic phase and study how the film thickness influences effective interactions.

%% file: theory.tex
\section{Theoretical background}

\subsection{The order tensor and the nematic director}
\label{order_tensor}
To form even the simplest liquid crystalline state one needs anisotropic molecules. To a good approximation, these highly anisotropic molecules can be thought of as rigid rods or ellipsoids of revolution. The orientation of every single molecule is then fully described by a unit vector ${\bm u}$ along its anisotropy axis.

Given the orientations of all the molecules ${\bm u}^{(i)}$ one can define an orientational order tensor
\begin{equation}
\label{order_parameter}
S_{\alpha \beta }\left( \bm{r}\right) =
\frac{3}{2N} \sum_{i}
\left[ u_{\alpha}^
{\left( i\right) }u_{\beta }^{\left( i\right) }-\frac{1}{3}\delta _{\alpha
\beta }\right],
\end{equation}
where the sum is over the all $N$ molecules in a small but macroscopic volume located at the point $\bm{r}$, $\bm{u}^{\left( i\right) }$ is a unit vector along the axis of the $i$-th molecule. The order tensor ${\bm S}$ is a symmetric traceless tensor and in general case has five independent components. Since the anisotropy of a macroscopic state originates from the molecular anisotropy on a microscopic level, the orientational order tensor Eq.~(\ref{order_parameter}) determines the
anisotropy of macroscopic properties of mesophase.

The director ${\bm n}$ is an eigenvector of the order tensor $S_{\alpha \beta}$ corresponding to the largest eigenvalue $S$. It gives an average orientation of the molecules in a mesoscopic volume located at the point ${\bm r}$. The order parameter $S$ has clear physical meaning: in the isotropic liquid $S = 0$, a perfectly aligned system has order parameter $S = 1$, and $S = -1/2$ if the molecules are randomly aligned in a plane.

\subsection{Frank-Oseen elastic free energy}
In the case where the scalar order parameter $S$ is constant, an expansion in terms of the director $\bm n$ is normally used to calculate elastic free energy density of a nematic
\cite{frank.fc:1958.a}
\begin{equation}
g_{\rm f}=
\frac{1}{2}\left\{
K_1\left( \nabla \cdot {\bm n}\right) ^{2}+
K_2\left({\bm n}\cdot \left[ \nabla \times {\bm n} \right] \right) ^{2}+
K_3\left[{\bm n}\times \left[ \nabla \times {\bm n} \right] \right] ^{2}
\right\},
\label{frank_free_energy}
\end{equation}
where $K_1$, $K_2$, and $K_3$ are Frank elastic constants representing splay, twist, and bend director distortions, respectively.

Equations for the director can be obtained by minimization of the free energy functional $\int_V g_{\rm f} d{\bf r}$ with an additional constraint ${\bm n}^2 = {1}$ and have the following form
\begin{eqnarray}
-K_1\nabla \left(\nabla \cdot {\bm n} \right) +
K_2\left\{ A [\nabla \times {\bm n}] + [\nabla \times(A {\bm n})] \right\} +
K_3\left\{ [[\nabla \times {\bm n}] \times {\bm B}] +
[\nabla \times[{\bm B} \times {\bm n}]] \right\}
+ \mu {\bm n}= 0,
\label{euler_lagrange_eqn}
\end{eqnarray}
where $A = {\bm n} \cdot [\nabla \times {\bm n}]$,
${\bm B} = \left[ {\bm n} \times [\nabla \times {\bm n} ]  \right]$,
$\mu$ is an unknown Lagrange multiplier which must be found from
the condition ${\bm n}^2 = {1}$.

\subsection{Landau-de Gennes free energy}
\label{free_energy}
Following the general approach to phase transitions given by Landau theory~\cite{landau.ld:1969.a} one can assume that the Gibbs free energy density $g\left( P,T,S_{\alpha \beta }\right) $ is an analytic function of the order parameter $S_{\alpha \beta }$. Expanding $g$ in a power series and taking into account invariance of $g$ under rotations one can obtain \cite{stephen.mj:1974.a}
\begin{equation}
\label{free_energy_lg}
g_{\rm bulk}=a\mathrm{Tr}{\bm S}^{2}-
b\mathrm{Tr}{\bm S}^{3}+
c\left[\mathrm{Tr}{\bm S}^{2}\right] ^{2},
\end{equation}
which is the most general expansion correct up to the fourth order in $S_{\alpha \beta}$. The coefficients $a$, $b$, and $c$ are in the general case functions of pressure $P$ and temperature $T$. In practice, $a$ is assumed to depend linearly on the temperature, $b$ and $c$ are considered temperature independent.

For this free energy, the \emph{uniaxial nematic} state is stable when $b^{2}>24ca$ with the degree of orientational ordering
\begin{equation}
S_{b}=\frac{b}{8c}\left( 1+\sqrt{1-\frac{64ca}{3b^{2}}}\right) .
\label{LDG_bulk_order}
\end{equation}

The order tensor ${\bm S}$ can depend on the spatial coordinates, which means that either the director ${\bm n}$ or the scalar order parameter S (or both) vary  from point to point. This variation can be due to external forces imposed on the system, thermal fluctuations, or boundary conditions. Assuming these deformations to vary slowly in space relative to the molecular distance scale, it is possible to describe the response of the liquid crystal using continuum elastic theory. Then the Gibbs free energy density, relative to the free energy density of the state with the uniform orientational order can be written as
\begin{equation}
\label{free_energy_nonuniform}
g_{\rm d}=L_{1}\frac{\partial S_{ij}}{\partial x_{k}}\frac{\partial S_{ij}}
{\partial x_{k}}+L_{2}\frac{\partial S_{ij}}{\partial x_{j}}\frac{\partial
S_{ik}}{\partial x_{k}}+L_{3}\frac{\partial S_{ij}}{\partial x_{k}}\frac{
\partial S_{ik}}{\partial x_{j}},
\label{LDG_gradients}
\end{equation}
where $L_1$, $L_2$, and $L_3$ are the elastic constants and the summation
convention is assumed.

\subsection{Density functional theory}
\label{onsager}
Onsager theory \cite{onsager.l:1949.a} is a form of a molecular field theory, with a distinctive feature that the configurational average energy is replaced by the orientation-dependent configurational entropy. It is a forerunner of the modern density-functional theory \cite{evans.r:1992.a}.

Consider an ensemble of elongated particles interacting pairwise through some potential ${\cal V}({\bf 1},{\bf 2})$ which depends on both positions and orientations of the molecules ${\bf 1}$ and ${\bf 2}$, where ${\bf 1} = ({\bm r}_1, {\bm \Omega}_1)$ and ${\bf 2} = ({\bm r}_2, {\bm \Omega}_2)$. Onsager has shown how the Mayer cluster theory may be used to give an
expansion for the equation of state of this system~\cite{onsager.l:1949.a}.
Onsager's expression for the Helmholtz free
energy per particle is expressed in terms of the single-particle
density, $\rho ({\bf 1})$, which gives the number of molecules
per unit solid angle and per unit volume
\begin{eqnarray}
\label{free_energy_onsager}
\beta F[\rho ]=\int \rho ({\bf 1})\left\{ \ln\rho ({\bf 1})\Lambda ^{3}-1-
\beta \mu + \beta U({\bf 1}) \right\} d({\bf 1})
-\frac{1}{2}\int
f({\bf 1},{\bf 2})\rho ({\bf 1})\rho ({\bf 2})d({\bf 1})d({\bf 2}).
\end{eqnarray}
Here $\beta = 1/k_{\rm B}T$ with $k_{\rm B}$ being the Boltzmann constant and $T$ the absolute temperature, $\Lambda$ is the de Broglie wavelength, $\mu$ is the chemical potential, $U({\bf 1})$ is the external potential energy,
\begin{equation}
f({\bf 1},{\bf 2}) = \exp \left[-{\cal V}({\bf 1},{\bf 2})/k_{\rm B}T\right]-1
\end{equation}
is the Mayer $f$-function, and  ${\cal V}({\bf 1},{\bf 2})$  is a pair-wise interaction potential.

The equilibrium single-particle density that minimizes the free energy functional in Eq.~(\ref{free_energy_onsager}) is the solution of the following Euler--Lagrange equation
\begin{equation}
 \ln\rho ({\bf 1})\Lambda ^{3}-\beta \mu + \beta U({\bf 1})-
\int f({\bf 1},{\bf 2})\rho ({\bf 2})d({\bf 2}) = 0,
\label{dens_onsager}
\end{equation}
which can be obtained from the variation of the functional
(\ref{free_energy_onsager}).

In principle,  Eq.~(\ref{dens_onsager}) can be solved numerically for any form of the second virial coefficient which can be expanded in Legendre polynomials. However, sometimes it is more convenient to minimize the functional in Eq.~(\ref{free_energy_onsager}) directly instead of solving the integral equation (\ref{dens_onsager}). The solutions are qualitatively the same as the solutions of the Maier-Saupe equation~\cite{stephen.mj:1974.a}. The difference is that the Maier-Saupe theory is commonly applied to liquids, which are only slightly compressible, whereas the Onsager expansion is applied to dilute suspensions of particles for which the change in free energy with density is relatively small. The Onsager theory predicts large changes in density at the transition, as is observed for such systems.
Note that the Onsager theory does not describe the bulk equation of state perfectly. The reason for this is the truncation of the virial expansion of the free energy at the leading, pairwise term. However, systematic improvements are possible which lead to the better agreement with the bulk equation of state
\cite{camp.pj:1996.b,parsons.jd:1979.a,lee.sd:1987.a,lee.sd:1989.a}.

\subsection{Monte Carlo and molecular dynamics simulations}
Computer simulations is a well established method to study liquid crystalline mesophases. It has been used to simulate various bulk properties of liquid crystals, such as bulk elastic constants~\cite{allen.mp:1988.a,allen.mp:1990.a,cleaver.dj:1991.a,phuong.n.h:2001.a,stelzer.j:1995.a,stelzer.j:1995.b,stelzer.j:1997.a}, viscosities~\cite{baalss.d:1986.a,sarman.s:1998.a}, helical twisting power \cite{allen.mp:1993.a,memmer.r:1998.a}, parameters of the isotropic-nematic interface \cite{allen.mp:2000.c}. Moreover, recent works have also proved that computer simulations can be effectively applied to study confined geometries~\cite{allen.mp:1999.a,andrienko.d:2000.c}.

Computer simulations can also be used to  study topological defects in liquid crystals. They provide a direct access to molecular positions and orientations and hence can directly validate aforementioned macroscopic and mesoscopic descriptions. In many cases, computer simulation provides information not extractable from phenomenological theories, such as the exact structure of the disclination line core \cite{andrienko.d:2000.b,hudson.sd:1993.a} or about the detailed liquid crystal structure around external inclusions \cite{andrienko.d:2002.a,Pereira2010}, defects structure and resulting effective interactions in nematic colloids \cite{Andrienko2003,Al-Barwani2004,Kim2004}, confirming the corresponding predictions of the continuous Landau-de Gennes theory \cite{Andrienko2003,Al-Barwani2004}.

%% file: single_particle.tex
\section{An isolated colloidal particle}

\begin{figure}
 \includegraphics[width=\textwidth]{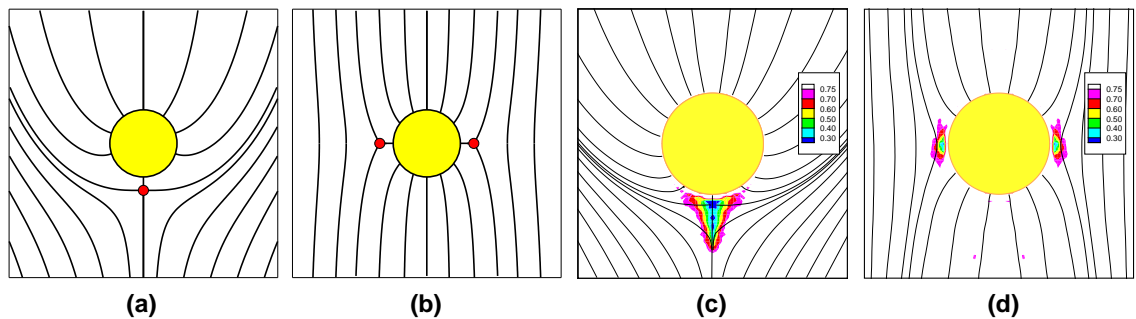}
 \caption[Sketches]{
 \label{fig:sketches}
(Color online) Sketches of (a) the satellite defect (dipole) and (b) the Saturn-ring defect (quadrupole) in the vicinity of a sphere with normal surface anchoring placed  in a uniform aligned nematic. (c) and (d) show the corresponding scalar order parameter profiles and director maps obtained from the molecular dynamics simulations. Adapted from Ref.~\cite{andrienko.d:2001.b}.
}
\end{figure}

In order to understand an effective interaction between particles dispersed in a nematic liquid crystal, it is essential to know the liquid crystal ordering near one of the particles. An isolated colloidal particle can provide either homeotropic (perpendicular to the surface) or tangential (parallel to it) boundary conditions for the nematic director. A particle with sufficiently strong homeotropic anchoring carries a topological charge of strength $+1$. If the director field is uniform far away from the particle, i.~e. the total charge of the whole system is zero, an additional defect in the medium must be created in order to compensate the topological charge of the particle. Two defect types can arise in this case. The first one is a hyperbolic hedgehog, a point defect characterized by the topological charge of $-1$, called  {\em satellite} defect. This particle-defect pair is usually modeled as a topological dipole  because the asymptotic behavior of the director field has dipolar symmetry. The other possibility is a quadrupolar, or a {\em Saturn-ring} defect, that is a $-1/2$ strength disclination ring that encircles the particle. Both are sketched in Fig.~\ref{fig:sketches}.

Theoretical and numerical work based on the Frank-Oseen elastic free energy~\cite{ruhwandl.rw:1997.b,lubensky.tc.1998,stark.h:1999,grollau.s:2003.a} as well as experimental observations~\cite{abbott.nl.2000} suggest that the dipolar configuration is stable for the micron-sized particles usually considered experimentally. The Saturn-ring configuration should appear if the particle size is reduced and, when present, is always predicted to be most stable in the equatorial plane normal to the far field director.  If the strength of the surface anchoring is weak enough, a surface-ring director configuration is possible~\cite{kuksenok.ov:1996.a,stark.h:1999}. It has also been found that a twisted dipolar configuration is possible for high enough ratios of splay to twist  elastic constants~\cite{stark.h:1999}.

In order to describe the defect structure, the elastic theory based on either Frank-Oseen, Eq.~(\ref{frank_free_energy}), or Landau-de Gennes, Eqs.~(\ref{free_energy_lg}), (\ref{LDG_gradients}), expressions for the free energy can be used. Note that the Frank-Oseen elastic theory has several restrictions: the defect core region cannot be treated and the liquid crystalline phase is considered as uniaxial. It has been demonstrated~\cite{biscari.p:1997.a,mottram.nj:1997.a,sigillo.i:1998.a,andrienko.d:2000.b,andrienko.d:2002.a} that the defect core of the simplest disclination line of strength $+1$ has a rather complex structure: the order parameter varies near the core region, the uniaxial phase becomes biaxial, and even the density of the mesophase can be affected by the presence of a wall or a defect~\cite{andrienko.d:2001.b,allen.mp:1999.a}. Hence, for a better description, it is necessary to use the tensorial order parameter formalism~\cite{sonnet.a:1995.a,schopohl.n:1987.a,schopohl.n:1988.a,fukuda:2002,fukuda:2004.a}.  This description has no anomalous behavior at the singularity of the director field and can resolve order parameter variations as well as biaxiality of the defect core. In addition, density variations can be taken into account using Onsager theory~\cite{allen.mp:1999.a}.

%% file: pair_interactions.tex
\section{Effective pair interactions}

Several types of interactions exist between the particles in a colloidal suspension: van der Waals, screened Coulombic, steric, fluctuation-induced Casimir forces, depletion, etc. In a nematic solvent, elastic deformations of the director around particles lead to an additional, long-range, interaction. It can be of a dipolar or quadrupolar type, depending on the symmetry of the director configuration around particles.

The first experimental observations of these interactions were made in inverted nematic emulsions, that is water droplets dispersed in a nematic solvent~\cite{poulin:1997.a}. The colloidal droplets formed linear chains, which were breaking upon transition to the isotropic phase. It was concluded that the  liquid crystal host induces dipolar interaction which leads to droplet chaining, as well as a shorter range repulsive interaction~\cite{stark.h:2001}.
Later, both dipolar and quadrupolar colloidal forces were measured directly using ferrofluid droplets in a magnetic field~\cite{poulin:1997.b}, magneto-optic tweezers~\cite{kotar:2006.a,kotar:2006.b,vilfan:2008}, iron particles in a magnetic field~\cite{noel:2006}, and dual-beam optical tweezers~\cite{kishita:2010,takahashi.k:2008.a,takahashi.k:2008.b}.

We will first discuss the asymptotic behavior (large inter-particle separations) of forces mediated by a nematic host. Possible corrections at small separations due to reorientation of defects and higher order terms in the expansion of the elastic free energy are presented at the end of this section.

\subsection {Long-range asymptotic behavior}
The large distance behavior of nematic-mediated interactions can be obtained from an electrostatic analogy by assuming that all constants in the expression for the elastic free energy, Eq.~(\ref{frank_free_energy}), are equal (one-constant approximation, $K_1=K_2=K_3=K$) and by performing a multipole expansion of the director field~\cite{lubensky.tc.1998,ramaswamy:1996}. Indeed, far away from a colloidal particle, the director can be written in a linearized form, ${\bm n}({\bm r}) = (n_{x},n_{y}, 1-O(n_{x}^{2},n_{y}^{2}))$, and the Frank-Oseen elastic free energy, Eq.~(\ref{frank_free_energy}), reads as
\begin{equation}
F \simeq \frac{K}{2}\int d^3r \Bigl  ((\nabla n_{x})^2 + (\nabla n_{y})^2 + O(n_{x}^{4},n_{y}^{4})  \Bigr ).
\label{frank-ossen}
\end{equation}
Hence, the transverse director components, $n_x$ and $n_y$, fulfill the Laplace equation. For a single particle, these can be expanded into multipoles. Using the superposition approximation, one can derive the following expressions for the effective pair potential
\begin{eqnarray}
\label{eq:pair_int}
\label{multipolar_potentials}
V_{\rm dipole-dipole} &\propto&\frac{1-3\cos^2\theta}{d^3}, \nonumber \\
V_{\rm quadrupole-quadrupole} &\propto& \frac{9-90\cos^2\theta+105\cos^4\theta}{d^5}, \\
V_{\rm dipole-quadrupole} &\propto& \frac{\cos\theta}{d^4}(15\cos^2\theta-9), \nonumber
\end{eqnarray}
where $d$ is the distance between the particles and $\theta$ is the angle between the far field orientation of the nematic director (the $z$ axis in our case) and the vector connecting the centers of the colloidal particles. 
These expressions can be generalized for particles of arbitrary shapes~\cite{Lev1999,Lev2002,Lev2004} or for weak anchoring at a particle surface~\cite{ruhwandl.rw:1997.a}.

The asymptotic behavior of the effective interaction potentials, given by Eqs.~(\ref{multipolar_potentials}), has been directly measured for the dipole-dipole configuration corresponding to $\theta=0$, using the free-release and the dual-beam optical tweezing methods. For parallel dipoles, this force was found to be attractive and  decaying with the particle-particle separation as $d^{-4}$~\cite{takahashi.k:2008.a,takahashi.k:2008.b,Skarabot2007,kishita:2010}. However, for the antiparallel orientation of dipoles, the pre-asymptotic decay was observed and the repulsive force decayed as $d^{-3.6}$~\cite{takahashi.k:2008.b}. Numerical minimization of the Landau-de Gennes free energy functional predicted an even slower decay of $d^{-3}$~\cite{Zhou2008,Fukuda2004}. In this work, the force was evaluated by numerically differentiating the Landau-de Gennes free energy. The direct integration of the corresponding stress tensor resulted in a decay of $d ^{-3.63}$~\cite{fukuda:2005.a}. Although the distance between the particles was as large as five particle diameters, the anti-parallel alignment was still in a pre-asymptotic regime. The dipole-quadrupole attraction with a decay of $d ^{-5}$, corresponding to $\theta = \pi$ in Eqs.~(\ref{multipolar_potentials}), was also confirmed experimentally~\cite{Ognysta2008}. The quadrupolar tail of $d^{-6}$ was also found experimentally for colloidal particles with the tangential boundary conditions~\cite{kotar:2006.a,Smalyukh2005}. Theoretically, the quadrupole-quadrupole asymptotic behavior was studied in detail using molecular theory, which accounts for the anisotropy of the molecular interactions~\cite{Sokolovska2008}.

\subsection {Short-range behavior}

The electrostatic analogy of the linearized Frank-Oseen free energy in the one-elastic-constant approximation, Eq.~(\ref{frank-ossen}), gives a good physical insight to understanding of structural properties of particle assemblies in dipolar~\cite{Skarabot2007}, quadrupolar~\cite{Skarabot2008}, or mixed~\cite{Ognysta2008,Ognysta2009} situations. For example, the formation of long chains in the first, or rhomboidal structures in the last two cases can be inferred from the $\theta$-angle dependence of the potentials in Eqs.~(\ref{multipolar_potentials}).

\begin{figure}
\includegraphics[width=\textwidth]{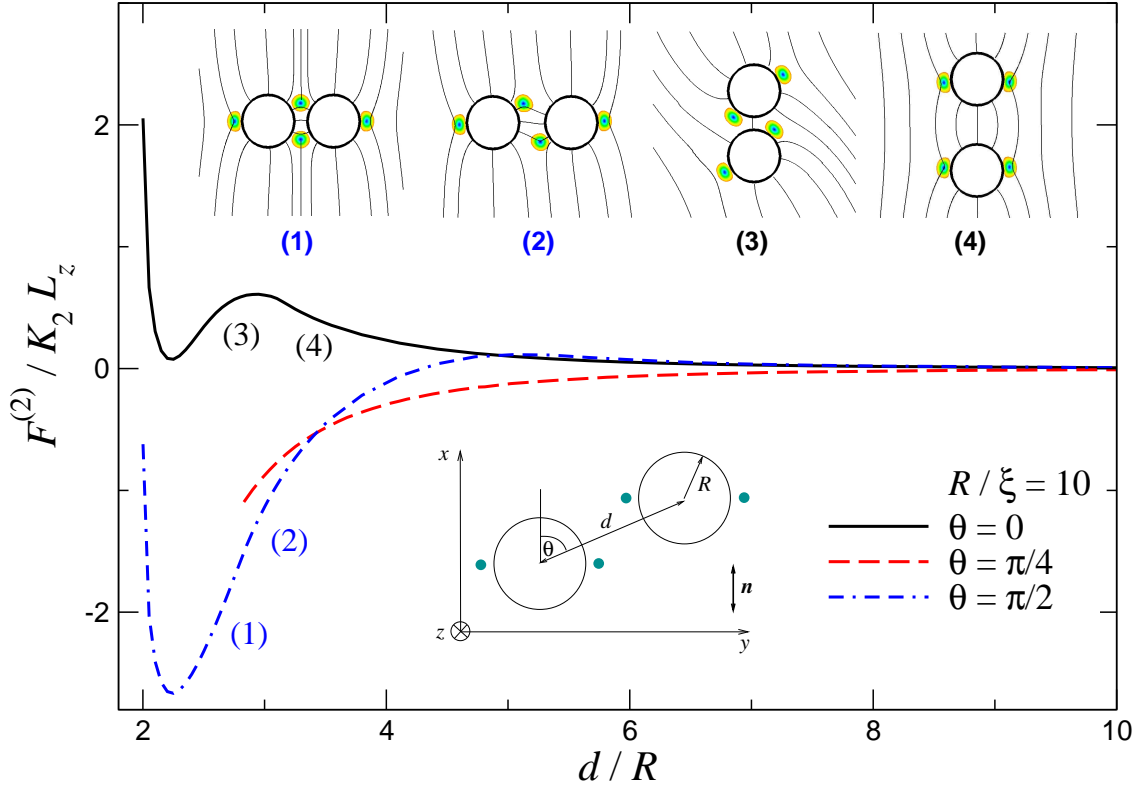}
\caption[]{(Color online) Effective interaction potential (or potential of mean force) as a function of the separation $d$ between two long cylindrical rods aligned along the $z$ axis. Several angles $\theta$ between the far field director and the center-to-center vector are shown. The lower inset shows the $z$ cross-section. The free energy is calculated by numerically minimizing the Landau-de Gennes free energy functional, Eqs.~(\ref{free_energy_lg}), (\ref{LDG_gradients}). The far field director is along the $x$ axis. The particle surfaces provide rigid homeotropic anchoring  which gives rise to two disclination lines of topological charge of $-1/2$ at each particle. The alignment of the director and the variation of the scalar order parameter are shown in the top inset. The typical values for a nematic compound 5CB~\cite{KRALJ1991} are used in Eqs.~(\ref{free_energy_lg}), (\ref{LDG_gradients}): $a = 0.06 \times 10^6 \,{\rm J/m^3}$, $b = 0.816 \times 10^6 \,{\rm J/m^3}$, $c = 0.45 \times 10^6 \,{\rm J/m^3}$, $L_1 = 6 \times 10^{-12} \,{\rm J/m}$,  $L_2 = 12 \times 10^{-12} \,{\rm J/m}$, and  $L_3 = 0$. $K_2=9S_\textrm{bulk}^2L_1/2$. The nematic coherence length  $\xi = \left( 48 L_1 c/b^2 \right)^{1/2} \approx 10\, \rm nm $, which corresponds to 5CB at a nematic-isotropic coexistence. The radius of colloidal particles is $R/\xi = 10$. The system size is $L_x\times L_y = 40R\times 40R$.
 \label{fig:2D-two_discs}
}
\end{figure}

\begin{figure}
\includegraphics[width=\textwidth]{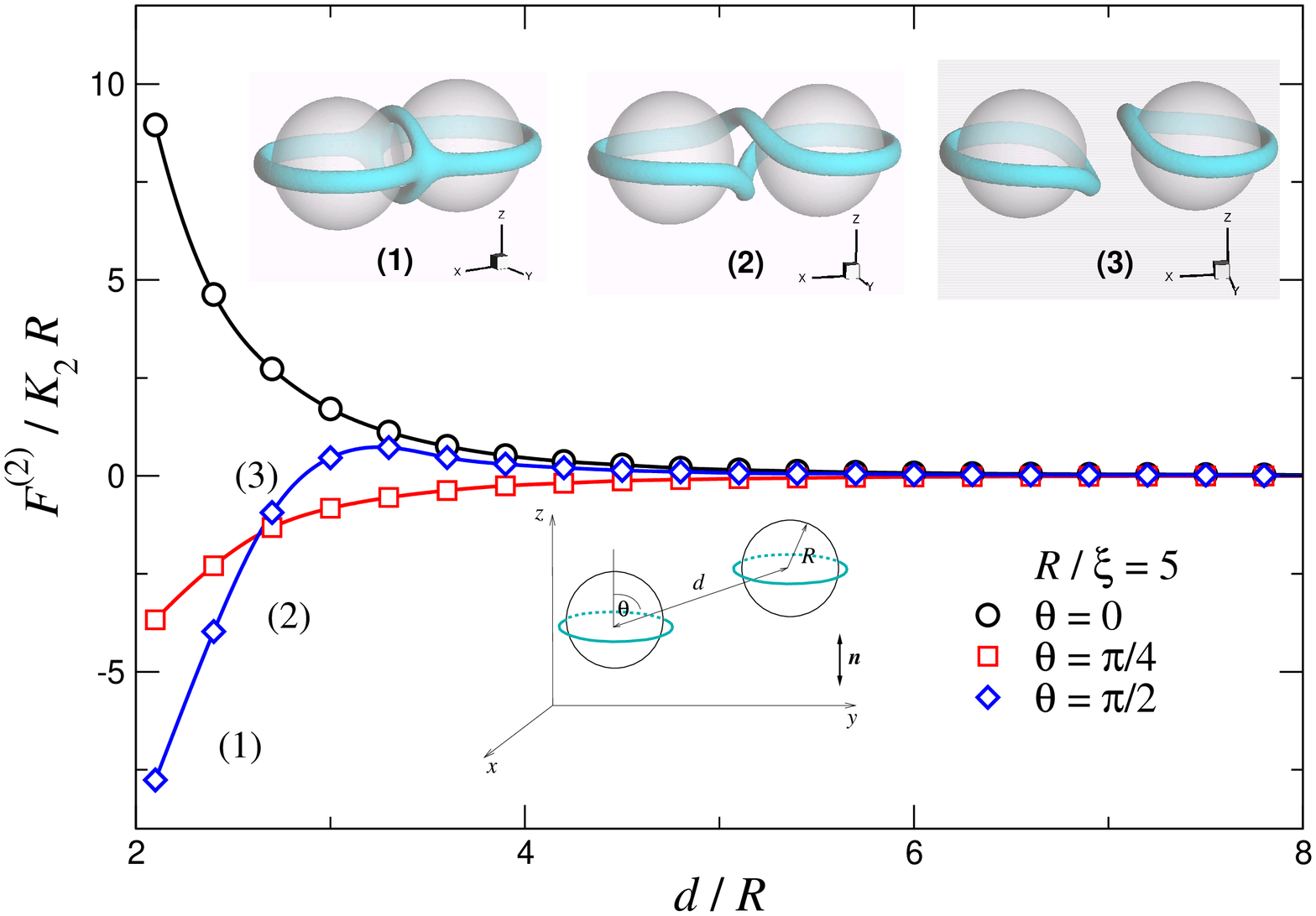}
\caption[Spheres]{(Color online) Effective interaction potential between two spherical colloidal particles as a function of separation $d$. The geometry is sketched in the lower inset. As before, the Landau de-Gennes free energy functional, Eqs.~(\ref{free_energy_lg}), (\ref{LDG_gradients}), is minimized numerically using the finite elements method and adaptive meshes. Particles provide rigid homeotropic boundary conditions for the director. The surface scalar order parameter is fixed to the bulk value of the order parameter, $S_{\rm surface} = S_{\rm bulk}$.  Material parameters used in Eqs.~(\ref{free_energy_lg}), (\ref{LDG_gradients})  are similar to those used for cylindrical particles, see caption to Fig.~\ref{fig:2D-two_discs}, except that $a = 0.01 \times 10^6 \,{\rm J/m^3}$. The upper inset depicts the nematic configurations for $\theta = \pi/2$ and several values of $d$. The blue rings  around the particles show the iso-surfaces of the scalar order parameter corresponding to $S=S_{\rm bulk}/2$. The radius of colloidal particles is $R/\xi = 5$. The system size is $L_x\times L_y\times L_z = 30R\times 30R\times 30R$.
\label{fig:two_spheres}
}
\end{figure}

However, at small separations, the situation becomes significantly more involved, since displacements of and interactions between topological defects lead to non-linear and many-body terms in the effective interaction potential. These short-range forces can be used to ``synthesize colloidal molecules''~\cite{Ognysta2009} through a direct assembly using optical tweezers. Such aggregates are relatively stable objects made of dipolar and quadrupolar ``colloidal atoms'' which are bonded by topological defects. To obtain the free energy associated with a ``chemical bond'', one needs to refine the description of the nematic mesophase and to take into account the behavior of topological defects at small separations.

Several approaches which go beyond the linearized theory and a simple superposition approximation have been proposed. Most popular are (i) numerical minimization of the Frank-Oseen elastic free energy functional in two and three spatial dimensions~\cite{stark.h:1999,Patricio2002,Korolev2008} (ii) minimization of the Landau-de Gennes free energy functional based on the five component tensorial order parameter~\cite{Zhou2008,Fukuda2004,fukuda:2005.a,Tasinkevych2002,Guzman2003,Silvestre2004,fukuda:2005.b,Hung2009,Ravnik2009},  (iii) molecular dynamics \cite{Andrienko2003} and Monte Carlo simulations \cite{ruhwandl.rw:1997.b}, or using classical density functional theory \cite{Cheung2007,Cheung2008}. The results of these efforts can be summarized as follows. At small separations between particles the topological defects rearrange their spatial position in order to minimize the free energy. This structural rearrangement changes the character of the effective interactions as compared to the asymptotic multipolar expressions in Eqs.~(\ref{multipolar_potentials}). These changes are illustrated in Fig.~\ref{fig:2D-two_discs} for the case of two long cylindrical and in Fig.~\ref{fig:two_spheres} for two spherical colloidal particles. The mismatch in the boundary conditions at the surfaces of the particles and the bounding box produces compensating topological defects:  accompanying disclination lines in the case of cylinders, and equatorial rings in the case of spheres. The spatial arrangement of these defects depends on the distance between the particles.
In both cases, the far-field director distribution has quadrupolar symmetry, which results in $d^{-4}$ and $d^{-5}$ decay of the interaction potential for cylinders and spheres, respectively. However, at some separation, which depends on the angle $\theta$ and the particle type, repulsion between the particles changes to attraction. Switching between repulsion and attraction is more pronounced for cylinders than spheres, where it appears only for $\theta = \pi/2$. The insets of Figs.~\ref{fig:2D-two_discs}, \ref{fig:two_spheres} show that such behavior is due to the rearrangement of topological defects. For spherical particles, for example, two Saturn rings (equatorial defects) merge into one at small separations. 

%% file: SmC_films.tex
\section{Confinement effects and thin films}

One more example of a colloidal emulsion/suspension with complex anisotropic interactions between inclusions is a system with a free standing smectic film serving as the host liquid. Here, the effective interactions can be controlled not only by anchoring conditions at surfaces of droplets or particles, but also by the symmetry of the mesophase, e.~g. by using different smectic phases: SmA, SmC, or SmC*. Additionally, temperature, polarization, film thickness and droplet size can be used to tune these interactions~\cite{Dolganov2006a,Dolganov2007,Dolganov2009}. In the following section, we briefly summarize the results on the self-assembly of inclusions in free standing smectic films. A more comprehensive overview of this field can be found in Ref.~\cite{Bohley2008}.

\subsection{Inclusions in free standing smectic films}

Thin free-standing smectic-C (SmC) or chiral smectic-C* (SmC*) films are ideal hosts for studying the assembly of inclusions in two-dimensional anisotropic colloidal suspensions and dispersions. Indeed, there are several benefits in using thin films compared to bulk systems. First, a free-standing film is bounded by two liquid-air interfaces. This enhances the smectic order which may persist even at temperatures up to $30^{\circ}\, \unit{C}$ above the corresponding bulk SmC/SmC* - isotropic (or nematic, cholesteric) transition temperature and guarantees that the smectic ordering will not be destroyed upon heating of the film. Second, smectic ordering facilitates the interpretation of the optical film texture when observed between crossed polarizers. Finally, the free-standing film geometry allows one to use reflection as well as transmission optical modes for visual inspection of the films.

Depending on a liquid crystal compound used, different types of inclusions can be nucleated in a film upon heating: (i) isotropic droplets close to the SmC-isotropic transition~\cite{Voltz2004}, (ii) nematic (N) droplets nucleated in SmC films close to the bulk SmC-N transition~\cite{Dolganov2006a,Cluzeau2002,Cluzeau2002a,Dolganov2006}, (iii) cholesteric (N*) droplets nucleated in a SmC* film close to the corresponding bulk
phase transition~\cite{Dolganov2006a,Dolganov2007,Dolganov2009,Cluzeau2001,Cluzeau2003}, and (iv) smectic islands, or circular localized regions containing additional smectic layers~\cite{Dolganov2009,Dolganov2006,Pettey1998}.

Dispersions are normally prepared by using well-controlled cycles of heating/cooling of free standing smectic films~\cite{Cluzeau2003}. In order to initiate the nucleation of lower order inclusions, the smectic film is gradually heated  across the corresponding bulk transition temperature until the temperature of nucleation is reached. As soon as the nucleation process starts, the temperature is kept fixed. The process of nucleation depends on the film thickness. For thin films ($N_{layers} <24$), initially nucleated droplets continue to grow by coalescing, which stops when the diameter of inclusions reaches the size of $4-5\, \mu \unit{m}$. For thicker films, nucleated droplets rarely coalesce, but continue to grow until the equilibrium sets in within the film, and droplets have grown to the diameter of about $18-20\, \mu {\unit m}$~\cite{Cluzeau2003}. Hence, the concentration and the size of inclusions can be controlled by the temperature difference, speed of a heating cycle, and by the initial film thickness. Upon cooling down, the inclusions can be transformed into smectic islands i.~e. localized regions containing additional smectic layers~\cite{Dolganov2006}.

\begin{figure}
\centerline{\includegraphics[width=\textwidth]{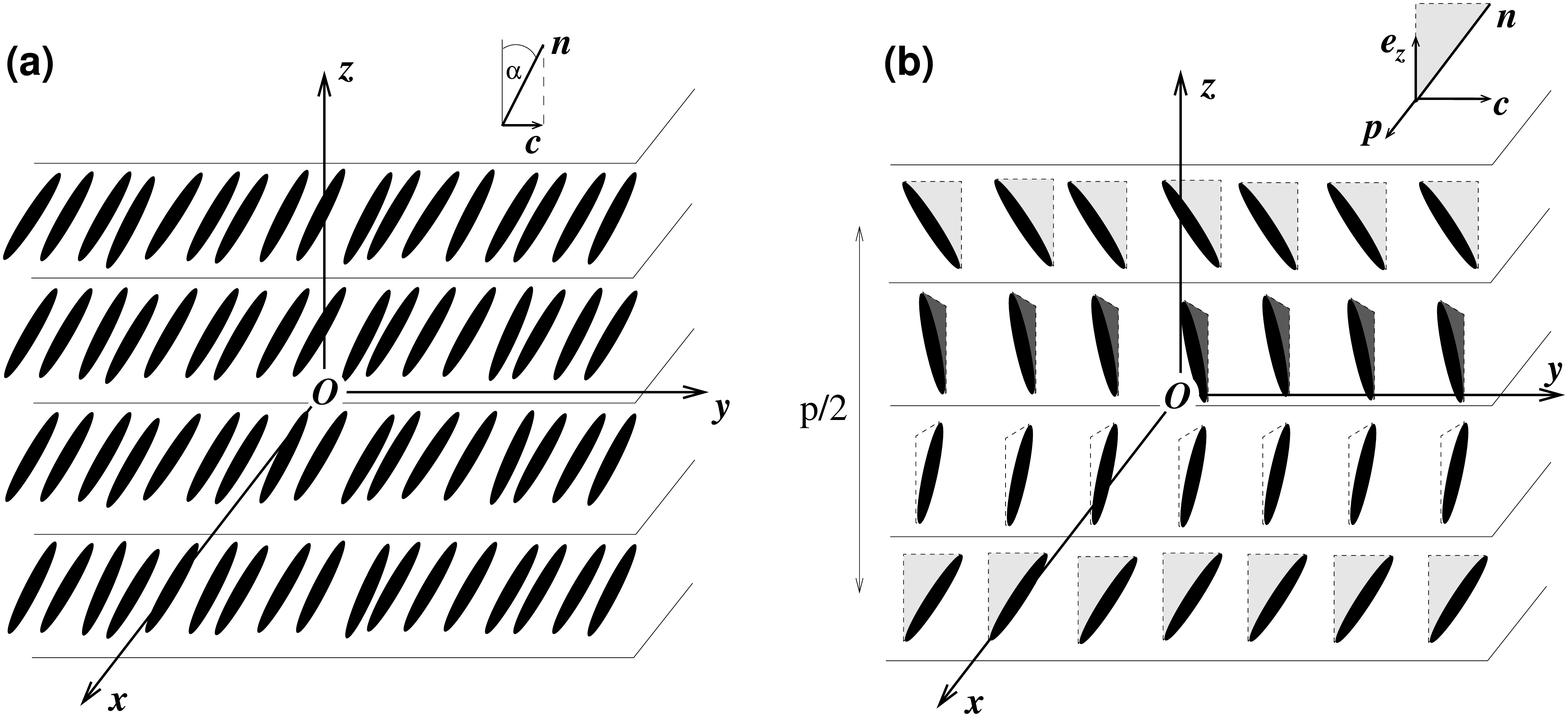}}
 \caption[]{(a) SmC and (b) SmC* phases. In the SmC phase the long molecular axis (or $\bm n$-director) is tilted with respect to the layer normal at an angle $\alpha$. SmC phase has monoclinic symmetry with the following group elements: (i) a plane of symmetry, the $zy$ plane, (ii) a two-fold rotation axis perpendicular to this plane ($x$-axis), and (iii) an inversion point $O$. A projection of the $\bm n$-director onto the plane of smectic layers defines the $\bm c$-director, which is a two dimensional vector. The SmC* phase is a chiral analogue of the SmC phase and is characterized by a helical rotation of the $\bm c$-director.  Due to the presence of the chiral molecules the $\bm c$-director twists as one progresses through the layers. The pitch $p$ is usually much larger then the inter-layer spacing. Chirality eliminates the symmetry plane as well as the inversion center and the remaining two-fold symmetry axis leads to a spontaneous polarization in its direction i.~e. SmC* is a ferroelectric material~\cite{MEYER1975}. The vector ${\bm p} = {\bm e}_z \times \bm{c}$ defines the polarization direction.
\label{fig:SmC_Cstar}
}
\end{figure}

Effective interactions between inclusions in a film depend strongly on the type of a liquid crystal. In a SmA phase, the $\bm n$-director, which determines the average orientation of the long axis of molecules, is normal to the smectic layers. Hence, additional elastic deformations lead to long-range isotropic interactions between inclusions. In SmC phases, the $\bm n$-director is tilted with respect to the layer normal by some angle $\alpha$, as shown in Fig.~\ref{fig:SmC_Cstar}a. The projection of the $\bm n$-director onto the plane of smectic layers defines a two-dimensional vector, the $\bm c$-director. Due to this tilt, SmC films provide an anisotropic environment for nucleated inclusions. The situation is even more complicated in SmC* films~\cite{MEYER1975}. This phase has the $C_2$ symmetry with a two-fold rotation axis normal to the $\bm n$-director and parallel to the smectic layers. Therefore, if a polarization has a component along the $C_2$ axis, it can not be canceled by the two-fold rotations around the axis. This leads to a net polarization in the chiral SmC* phase in the $\bm p$-direction as shown in Fig.~\ref{fig:SmC_Cstar}b. The coupling of this polarization to the $\bm c$-director gives rise to a temperature-induced rearrangement of topological defects\footnote{
Note that the accompanying topological defects of the $\bm c$-director field are different from those of the $\bm n$-director. Indeed, the  $\bm c$-director, contrary to the $\bm n$-director, is a two-dimensional vector. Hence, its ground state manifold is a unit circle $S^1$, and the fundamental  group is $Z$ (integers). Therefore, in free standing smectic films, the type of a topological defect generated by inclusions with homeotropic or planar boundary conditions is either a bulk or surface point defect of strength of $-1$ or two  surface defects of strength of $-1/2$.} and can even change  anchoring conditions at the boundaries of N* inclusions in free-standing ferroelectric SmC* films~\cite{Dolganov2007,Dolganov2009}.

Smectic-inclusion interfaces impose specific anchoring conditions on the $\bm c$-director. Isotropic or nematic inclusions in SmC films impose planar anchoring on the $\bm c$-director~\cite{Voltz2004,Cluzeau2002,Cluzeau2002a,Cluzeau2005,Dolganov2006}. A similar situation is observed for smectic islands~\cite{Dolganov2006,Pettey1998}. Cholesteric droplets nucleated in SmC* films impose homeotropic anchoring on the $\bm c$-director~\cite{Cluzeau2001,Cluzeau2003}, which can be rationalized by taking into account the spontaneous polarization of SmC*, which gives rise to a polarization-dependent term in the anchoring energy~\cite{Cluzeau2004}. This term favors a planar orientation of the polarization or a homeotropic orientation of the $\bm c$-director at the inclusion boundary, in order to achieve zero surface electric charge and to minimize the electrostatic energy of the inclusion.

In some cases, anchoring conditions at the inclusion boundary can depend on temperature and thickness of the SmC* film or even a surface transition can occur~\cite{Dolganov2007}. For example, during the nucleation process, just slightly above the bulk SmC*-N* transition, the $\bm c$-director has planar anchoring. Upon further heating, a series of transitions from planar to homeotropic anchoring and back is observed until the film thickness reaches a critical value of $N \approx 9$, below which only planar anchoring is possible. The transition from planar to homeotropic anchoring is initiated by splitting of the $-1/2$ surface defects into pairs of the $-1/4$ surface defects, which repel each other and slide along the inclusion boundary. Eventually, these intermediate defects of strength of $-1/4$ recombine either with the original intact $-1/2$ defect or with each other.

\begin{figure}
\includegraphics[width=\textwidth]{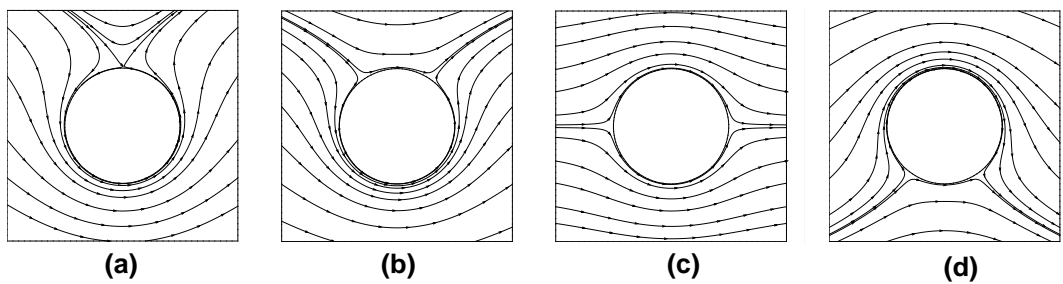}
\caption[]{Sketch of the $\bm c$-director streamlines around a circular inclusion in a two-dimensional SmC* film. Planar anchoring is assumed at the boundary of the inclusion. The dipole-quadrupole-reversed dipole configurational transition is illustrated: (a) a single surface defect of strength of $-1$ forming a dipolar configuration of the $\bm c$-director, (b) two surface defects of strength of $-1/2$ forming a mixed configuration with the decreased dipolar moment, (c) a quadrupolar configuration, and (d) a reverse mixed configuration made of two surface defects of strength of $-1/2$. This transformation can be induced by temperature variation and is reversible upon heating and cooling of the film, assuming the temperature does not cross the thinning transition as the number of smectic layers decreases by one~\cite{Dolganov2007,Dolganov2009}.
 \label{fig:SmC_schematic}
}
\end{figure}

A change in polarity of thin smectic films can also lead to a change of the type of defects around the inclusions. For example, in a nonpolar membrane of SmC, the configuration of the $\bm c$-director field is quadrupolar with two $-1/2$ surface defects, while it is dipolar with one $-1$ surface defect in the case of polar membrane of SmC*~\cite{Dolganov2006a}. A similar transformation between the dipolar and quadrupolar configurations can be performed reversibly by changing the temperature of the sample~\cite{Dolganov2007,Dolganov2009}. The transformation of the  $\bm c$-director proceeds through splitting of the surface topological defect of strength of $-1$ into two surface defects of strength of $-1/2$ (see Fig.~\ref{fig:SmC_schematic}). As the temperature changes, the $-1/2$ defects move along the boundary and may recombine on the opposite side of the inclusion resulting in the reorientation of the $\bm c$-director at the boundary by an angle equal $\pi$.

Anchoring conditions determine the type and the topological charge of defects accompanying inclusions and therefore define their self-assembly behavior.
For example, cholesteric droplets in SmC* films can provide homeotropic anchoring for the director. In this case, a formation of chains is observed which is due to the effective dipole-dipole interactions between the inclusions~\cite{Cluzeau2001,Cluzeau2003}. On the other hand, planar anchoring is observed for nematic inclusions in SmC films~\cite{Cluzeau2002,Cluzeau2002a}. In this case, the $\bm c$-director field around each inclusion has a quadrupolar symmetry, which leads to the formation of kinked chains at low concentrations of inclusions and to two-dimensional structures with a square lattice at higher concentrations.
In some situations, inclusions in SmC* films impose planar anchoring. In this case, for dipolar inclusions, chains orient perpendicular to the far field $\bm c$-director. For a quadrupolar configuration, kinked chains and rhomboidal clusters of four particles are formed~\cite{Dolganov2006a,Dolganov2009}.

In spite of a significant amount of experimental work, theoretical understanding of these systems is rather limited. The main reason is the complexity of their description, which often requires techniques suitable to treat several length scales, from smectic layering to long-range distortions of the director field around inclusions. In addition, inclusions have soft boundaries, which pose additional problems when specifying boundary conditions.

The effective elastic force between particles in free-standing SmC films was shown to have an asymptotic power-law of $d^{-1}$ decay, where $d$ is the  separation between the particles, and to be inversely proportional to the smectic-air surface tension $\gamma$~\cite{deoliveira:2009}. For small separations, a superposition approximation employing the analytical single-particle solution was used to calculate the effective interaction between inclusions with planar anchoring~\cite{Silvestre2008}. For weak anchoring conditions, the global minimum of the free energy corresponds to a configuration where inclusions are in a contact and the center to center vector forms an angle of $\approx 36^{\circ}$ with the far field $\bm c$-director. For anchoring strengths above a threshold value, there is an optimal separation between inclusions which increases with the anchoring strength and can reach a value of three inclusion diameters. Additionally, a pair of inclusions becomes more aligned with the far field $\bm c$-director as the anchoring increases.

The case of chiral SmC*  can be modeled by introducing an additional elastic term, $\propto \nabla \times \mathbf{c}$, to the two-dimensional elastic free energy density of SmC~\cite{Bohley2007,Fukuda2007}. This term allows for a spontaneous bend deformation of the $\bm c$-director and favors the dipolar defect configuration with two $-1/2$  surface defects placed at one side of the inclusion. Without this term, the quadrupolar configuration of the $\bm c$-director with two antipodal surface defects of strength of $-1/2$ was found to be energetically more stable~\cite{Bohley2006,Cluzeau2005,Lejcek2005}.

The aforementioned work treats inclusions as if they were rigid bodies, which is clearly an oversimplification for liquid-like inclusions, since the anchoring as well as the surface tension is coupled to the director orientation at the inclusion surface and hence to the structure of the surrounding film. The latter varies with temperature, droplet size, external field, etc. The shape of the soft inclusion in the SmC film with planar anchoring was studied using the two-dimensional elastic free energy supplemented by two phenomenological terms describing the anchoring energy and the line tension respectively~\cite{Silvestre2006}. It was found that the anchoring term  favors  an elongated shape of inclusions, while the line tension term favors the circular one. For any finite anchoring strength there is a critical value of the line tension parameter at which the elliptical shape of the inclusion becomes unstable.

%% file: nematic_cells.tex
\subsection{Quasi two-dimensional nematic colloids}

\begin{figure}
\centerline{ \includegraphics[width=0.5\textwidth]{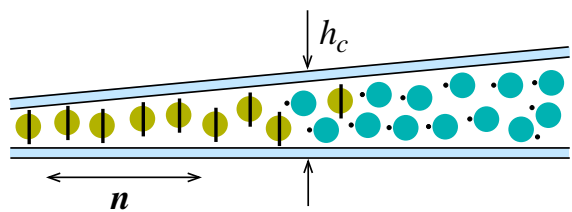}}
\caption[2D_nematic_geometry]{(Color online) Sketch of a quasi-two-dimensional nematic colloid sandwiched between two glass plates in a wedge-like geometry. The cell walls impose planar and the colloidal particles homeotropic anchoring on the director. In the thinner part of the cell colloidal particles generate quadrupolar configurations in the surrounding nematic field, while in the thicker part of the cell dipolar configurations are more stable. Hence, the dipole-quadrupole configurational transition is governed by the confinement. For $\approx 2.32 \mu \unit{m}$ silica particles dispersed in 5CB, the threshold cell thickness is $h_c\approx 3.5 \mu \unit{m}$~\cite{Musevic2006}.
\label{fig:2D-nematic_film}
}
\end{figure}
A different route to control the two-dimensional crystallization of colloids is to use a confinement-induced change of the charge of a topological defect accompanying a colloidal particle~\cite{Skarabot2007,Skarabot2008,Ognysta2008,Ognysta2009,Musevic2006}. To achieve this, a dispersion of particles is confined in a wedge-like cell. The surfaces of the cell provide planar anchoring, while particles impose homeotropic anchoring of the director, as shown in Fig.~\ref{fig:2D-nematic_film}. As it turns out, the Saturn ring defect (quadrupolar symmetry) is stable in the thinner part of the cell, while the satellite defect (dipolar symmetry) is favored in its thicker part. At a threshold thickness, $h_c/R \approx 3$, colloidal particles with both dipolar and quadrupolar director configurations can coexist with each other~\cite{Ognysta2009,Musevic2006}. Hence, the dipolar to quadrupolar transformation of the symmetry of the director field around the colloidal particles can be achieved by changing the cell thickness, not the particle size.

\begin{figure}
 \includegraphics[width=\textwidth]{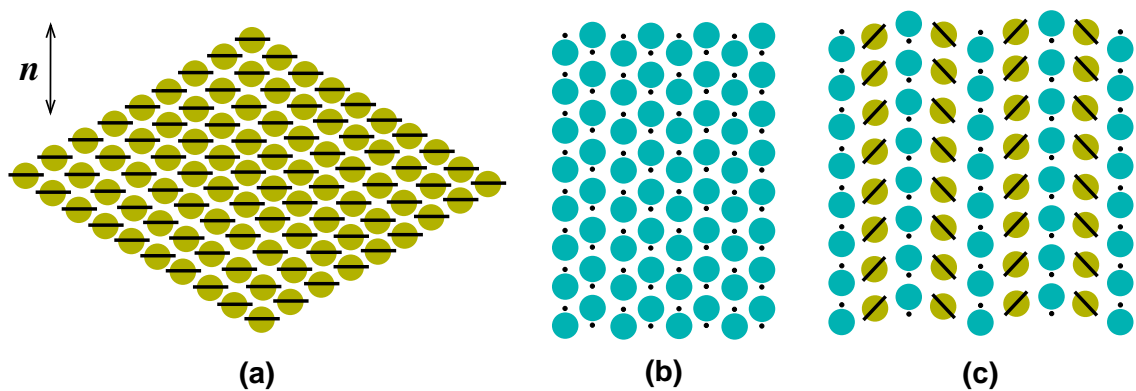}
 \caption[2D_colloidal crystallites]
{(Color online) Sketch of two-dimensional colloidal crystallites composed of (a) quadrupolar~\cite{Skarabot2008}, (b) dipolar~\cite{Skarabot2007}, and (c) mixed~\cite{Ognysta2008} colloidal particles. The crystallites are assembled with the help of laser tweezers.
 \label{fig:2D-crystallites}
}
\end{figure}

For thin enough cells, where only quadrupolar colloidal particles are present, particles self-assemble into long kinked chains and only sometimes small crystallites of nine and less particles appear~\cite{Skarabot2008}. Bigger crystals can be assembled by using laser tweezers. The resulting crystal structure is shown in Fig.~\ref{fig:2D-crystallites}a. The measured strength of the ``bonds'' in such a crystal is rather weak compared to the case of dipolar colloids. For example, when a particle approaches the head of a kinked chain, the measured binding energy is of the order of $10^3 k_{\rm B} T$ for $2.32 \mu \unit{m}$ particles. For a lateral approach, the binding energy is of the order of $0.3\times 10^3 k_{\rm B} T$. Due to the anisotropy and weakness of the binding energy, the assembled crystallites are rather fragile and ``melt'' even for small variations of temperature.


Contrary to quadrupolar crystallites, the ones assembled from colloids with dipolar defects were found to be much more stable~\cite{Skarabot2007}. The measured binding energy for aligned dipoles is of the order of $5\times 10^3 k_{\rm B} T$ for $2.32 \mu \unit{m}$ particles (in both directions, $\theta=0$ and $\pi/2$). The arrays of long dipolar chains, forming a 2D colloidal crystal, were directly assembled as shown in Fig.~\ref{fig:2D-crystallites}b. Each chain has a ``ferroelectric'' order of topological dipoles, and therefore the whole 2D assembly possesses ``antiferroelectric'' ordering. The effective interaction potential between two dipolar particles was also directly measured as a function of their separation $d$~\cite{Skarabot2007}. However, the accuracy of the measurements did not allow the comparison to the theoretical predictions, which concluded that the dipole-dipole power law of $d^{-3}$ is exponentially screened for distances larger than the thickness of the cell, with an extended crossover region between the two regimes~\cite{Chernyshuk2010}.


In the coexistence region, where both dipolar and quadrupolar configurations are stable, the corresponding trapping energy is of the order of $5\times 10^3 k_{\rm B} T$ for $4\, \mu \unit{m}$ particles~\cite{Ognysta2008,Ognysta2009}. Close to a dipolar particle with the fixed orientation of its topological dipole, three different equilibrium configurations of the second quadrupolar particle are possible. This allowed the assembly of a variety of two-dimensional colloidal crystals with different unit cells, one of these crystals is shown in Fig.~\ref{fig:2D-crystallites}c. The measured dipole-quadrupole pair interaction when particles approach each other along the rubbing direction ($\theta = 0$) with the hyperbolic hedgehog defect placed between them, agreed well with the asymptotic behavior of $d^{-5}$. No measurable effects of the cell thickness on the power-law exponent was found, however, the prefactor of the interaction energy was found to be thickness dependent.

The self-assembly idea was further extended to the field of ``colloidal chemistry''~\cite{Ognysta2009}, where colloidal particles play the role of atoms, which could, by sharing their topological defects, bind to ``colloidal molecules''. Several types of  such ``colloidal molecules'' were ``synthesized'' with the help of optical tweezers.  A number of exotic structures, such as one-dimensional ordered  wire-like structures were also assembled.  The particles in such a wire are bound by a common closed disclination line, with the binding energy  of the order of $10^4 k_{\rm B} T$ for $4.7 \mu \unit{m}$ particles, which is comparable to the binding energy along a dipolar chain~\cite{Ravnik2007}. In addition, an externally applied AC electric field was used to obtain close-packed hexagonal lattices~\cite{Skarabot2007a}.

%% file: interfaces.tex
\section{Colloids at interfaces}

Liquid-liquid interfaces provide an additional way of controlling colloidal assembly. In case of a two-dimensional organization of colloidal particles at fluid interfaces, interparticle forces are normally isotropic and lead to a hexagonal ordering of particles. Anisotropic interactions are also possible and can be observed for non-spherical particles trapped by fluid-fluid interfaces~\cite{Danov:2010}.

In anisotropic fluids, such as liquid crystals, a much richer variety of ordered patterns at interfaces can be observed, for example the chain-like assemblies of particles~\cite{smalyukh:2004.a,Lin2008,Koenig2010}. The type and degree of ordering can be controlled by adding a molecular surfactant. For example, reorganization of micro-particles was observed by the addition of an anionic surfactant SDS to the liquid crystal 5CB. First, the spacing between the particles within chains increased and ultimately two-dimensional arrays with local hexagonal symmetry were formed~\cite{Koenig2010}. These experiments demonstrated that reversible, chemosensitive control of the interfacial organization of colloidal particles is possible. An alternative method is to dope the emulsion with a photosensitive amphiphilic dye. Upon exposing the emulsion to ultraviolet irradiation, the dye undergoes trans-cis isomerization, which changes the nematic-air surface tension, anchoring on the surface of droplets, and elastic constants of the host. In this way one gains a control of the lattice constant of the two-dimensional hexagonal crystal formed by the glycerol droplets at the nematic-air interface~\cite{Lev2008}. By irradiating the emulsion with spatially-modulated light one may even obtain patterned two-dimensional colloidal crystals. More exotic situations include coexistence of different two-dimensional assemblies of glycerol droplets at a nematic-air interface, which can be controlled by reorientation of elastic dipoles around each droplet~\cite{Nych2007}.

Apart from controlling the assembly of particles at interfaces, it is also possible to use nematic-isotropic (NI) interfaces for ordering colloids into three-dimensional structures. For example, one can obtain cellular solids by dispersing colloidal particles in the isotropic phase of a liquid crystal and then quenching it below the NI transition temperature~\cite{Anderson2001,roth:2010}. Colloidal particles, densely packed at thin interfaces between different nematic domains, form an open cellular structure with a characteristic domain size of $10-100 \mu \unit{m}$. The size of the domains can be controlled by changing the particle concentration. Due to the cellular morphology, such gel-like solids have a remarkably high elastic modulus, which varies linearly with the particle concentration~\cite{Anderson2001a}.

The structural organization of colloidal particles can also be controlled by changing the speed of a moving NI interface. When a colloidal particle is captured by a NI interface, it can be dragged by it. As a result, periodic, stripe-like structures can be obtained, with the period depending on particle mass, size, and interface velocity~\cite{West2002,West2004a,West2006}. In combination with externally applied electric field, a wide variety of particle structures is observed, ranging from a fine-grained cellular structure to stripes of varying periods, to a coarse-grained ``root'' structure~\cite{West2004}. A combination of a moving nematic-isotropic interface with a patterned electric field can also be used to move particles from one place to another and pack them in a certain place in an ordered fashion. The speed of the interface, the magnitude of the applied electric field, particle size, density and its dielectric properties control which particles can be moved and which are left behind. This provides the possibility, for example, to place ``defects'' at particular locations in photonic crystals~\cite{West2005}.

From the theoretical standpoint, one can think of two origins of anisotropic interactions between particles pinned at an interface. First, the director deformations extend into the nematic phase, i.~e. in addition to interfacial energies, the bulk elasticity contributes to the total surface free energy of the system. Second, the effective surface tension depends on the orientation of the director in its vicinity and hence varies along the interface due to the presence of
colloidal particles.

Long-range elastic effects significantly complicate theoretical understanding of interparticle interactions at NI interfaces: expressions for the free energy or force-distance profiles can hardly be obtained analytically, especially for small separations, the case when also many-body effects are important. Asymptotic analysis predicts that, for the case of the homeotropic anchoring at the interface, the elastic interaction is quadrupolar, repulsive, and decays with the distance as $(R/d)^{-5}$~\cite{Oettel2008,Oettel2009}. Numerical calculations are also not straightforward because the interfacial width and the size of the defect core are normally much smaller that the size of a colloidal particle, hence there are several rather different length-scales involved. Therefore, significant computational efforts are required in order to address either static or dynamic problems. The remedy to this problem is either to solve the discretized Euler-Lagrange equations or to directly minimize the free energy functional using finite elements with adaptive meshes~\cite{Andrienko2004,Andrienko2005,tasinkevych:2006.a}. In what follows, we illustrate this approach by studying interaction of particles captured by a nematic-isotropic interface.

%% file: results.tex
\subsection{Example: Colloidal particles at a nematic-isotropic interface}

\begin{figure}
\centerline{\includegraphics[width=0.5\textwidth]{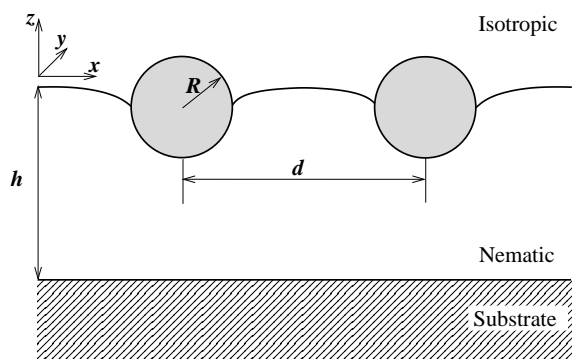}}
\caption[]{Schematic drawing of the geometry of the system considered. Nematic film of thickness $h$ is sandwiched between the isotropic phase and a solid substrate, imposing unidirectional strong planar anchoring on the director. Weak homeotropic anchoring is assumed at the NI interface. A small temperature gradient is applied in the $z$ direction, such that the free NI interface (without the particles) is stabilized at the $z=0$ plane (nematic is the thermodynamically stable phase for $z<0$). Colloidal particles of radius $R$ which are distance $d$ apart are trapped at the interface. We consider both strong homeotropic and planar degenerate anchoring at particle surfaces. In the case of a single colloidal particle, we vary the $z$-coordinate of the center of the particle. For a pair of particles, the $z$-coordinates of their centers are fixed at $z = 0$, and the separation $d$ is varied along the $x$ direction.
\label{fig:NI_interface_colloids}
}
\end{figure}

In this section, we illustrate how to determine the force on a colloidal particle from the side of a NI interface, as well as evaluate the pair force between two colloidal particles pinned at this interface. In order to avoid the divergence of the elastic free energy in the defect core, we use the expression for the free energy density in the Landau-de Gennes form, Eqs.~(\ref{free_energy_lg}), (\ref{LDG_gradients}). This free energy has to be minimized either under constraints imposed by the inclusion (colloidal particle) or by adding an appropriate surface term to the total free energy. A similar two-dimensional problem has already been treated in Refs.~\cite{Andrienko2004,Andrienko2005}. In what follows, we consider the three-dimensional case, with a flat unperturbed NI interface and spherical colloids\footnote{Note that the interaction between long rods considered previously in Refs.~\cite{Andrienko2003,Andrienko2004,Andrienko2005,tasinkevych:2006.a} differs quantitatively from the interaction between spherical particles because another type of topological defects accompanies spherical colloids: either a point defect of strength -1 or a ring defect of strength -1/2 (contrary to -1/2 strength line defects for cylinders).}.
The geometry is sketched in Fig.~\ref{fig:NI_interface_colloids} and is similar to the one studied experimentally in Refs.~\cite{smalyukh:2004.a,Nych2007}. We assume hybrid boundary conditions: a strong parallel anchoring at the substrate and a weak homeotropic one at the NI interface. The latter is tuned by choosing the appropriate anisotropy of elastic constants, $L_2/L_1 = -1/2$~\cite{DEGENNES1971}. We first calculate the force on a single particle as a function of its immersion in the nematic film. Two types of anchoring at the particle surfaces, homeotropic and planar, are considered, as well as different thicknesses of the nematic film. The planar degenerate anchoring at particle surfaces is modeled by the expression proposed in Ref.~\cite{Fournier2005}. We then proceed by calculating the effective interaction potential between two particles pinned at the interface and discuss bridging forces and defect configurations around the colloidal particles. To summarize, we compare our results with theoretical predictions at large separations as well as to experiments.

\subsubsection*{Single particle at a NI interface}
Let us first address the force on a single particle due to bending of a nematic-isotropic interface and elastic director deformations inside the nematic layer.
\begin{figure}
\centering
\includegraphics[width=\textwidth]{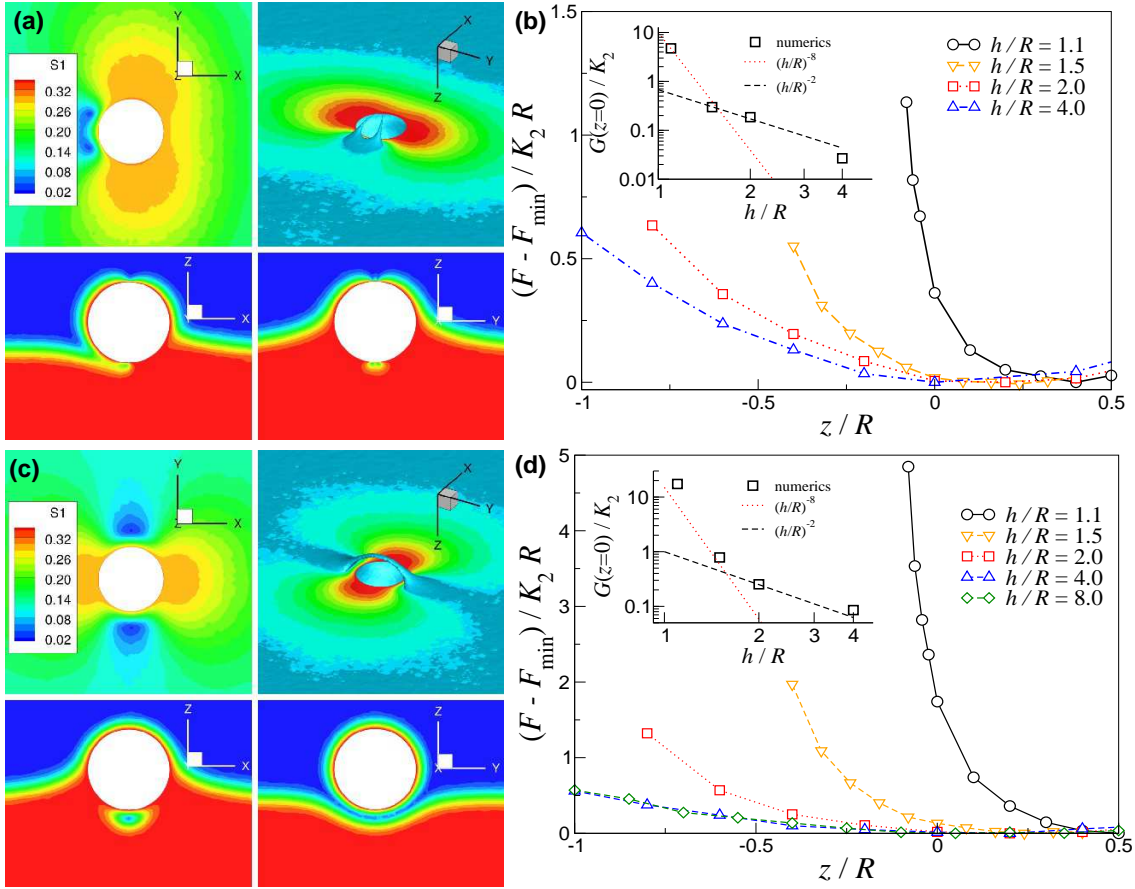}
\caption{Contour plots of the nematic order parameter for (a) planar and (c) homeotropic anchoring of the nematic director on the particle surface. The particle position is fixed at a distance $z/R=3/5$ above the NI interface. Plots (b) and (d) show the free energy as a function of particle immersion into the nematic film for planar and homeotropic anchoring, respectively. Several film thicknesses are shown. Insets illustrate the force on the particle evaluated by numerical differentiation of the corresponding free energy at $z=0$ as a function of the thickness of the nematic film. The twist elastic constant, $K_2=9S_{\rm bulk}^2L_1/2$, is evaluated at the bulk NI transition temperature. The following values for the material parameters are used in Eqs.~(\ref{free_energy_lg}), (\ref{LDG_gradients}):  $b = 0.816 \times 10^6 \,{\rm J/m^3}$, $c = 0.45 \times 10^6 \,{\rm J/m^3}$, $L_1 = 6 \times 10^{-12} \,{\rm J/m}$,  $L_2 = -3 \times 10^{-12} \,{\rm J/m}$, $L_3 = 0$, and $a=a_{NI}(1+z\times 0.004)$, where $a_{NI} = b^2/(24c)$  corresponds to the nematic-isotropic coexistence. In this case, $R/\xi = 5$, where $\xi$ is the nematic coherence length defined in the caption to Fig.~\ref{fig:2D-two_discs}. The system size is $L_x\times L_y = 30R\times 30R$.
\label{fig:single}
}
\end{figure}
The results of the minimization of the Landau-de Gennes free energy functional as given by Eqs.~(\ref{free_energy_lg}), (\ref{LDG_gradients}) are summarized in Fig.~\ref{fig:single}. Figs.~\ref{fig:single}a,c show several cross-sections of the scalar order parameter $S$, which is the largest eigenvalue of the order tensor $S_{\alpha \beta}$. As one can see, the structure of the topological defects is different for two types of anchoring. In  the case of homeotropic anchoring, a half Saturn ring defect is created in the nematic layer (the second half is merged into the isotropic phase), Fig.~\ref{fig:single}c, while for planar anchoring the defect structure is less symmetric and the defect core is split between two branches, each of them representing a disclination line which starts at the particle surface and ends in the isotropic phase. This defect cannot be easily identified with the standard  structures, such as a Saturn ring, a satellite, or a bojoom-like topological defects.

Different defect structures, however, do not lead to a quantitatively different dependence of the free energy on the particle immersion depth, which is shown in Fig.~\ref{fig:single}b,d. In both cases, the particle is expelled from the nematic phase with a force that depends on the films thickness $h$. For small immersion depths, the free energy is roughly  quadratic in the depth i.~e. the force on the particle is proportional to the depth. The expelling force is also larger for homeotropic anchoring, since this type of anchoring introduces more deformations into the nematic phase.

The asymptotic analysis for similar hybrid boundary conditions (both assumed to be rigid) predicted a power-law decay of the expelling force of at least $(R/h)^8$~\cite{Oettel2009}, which basically reflects the fact that stronger elastic deformations of the director field are created in thinner nematic films. It is clear from the insets of Fig.~\ref{fig:single} that this scaling does not hold in our case. The index of the exponent changes by practically six orders of magnitude. Note, however, that this might be an artifact of the assumption of the infinitely strong director anchoring at the particle surface, which results in stronger elastic deformations for thinner films. In addition, our calculations assume that the homeotropic anchoring at the NI interface is rather weak, contrary to infinitely strong homeotropic anchoring assumed in Ref.~\cite{Oettel2009}.

\subsubsection*{Interaction of two particles pinned at a NI interface}
We now look at the effective interaction potential of two particles pinned at a NI interface. We adopt the same geometry as for a single particle, with both particles fixed at $z=0$ and a separation $d$. The corresponding geometry is sketched in Fig.~\ref{fig:NI_interface_colloids}.

\begin{figure}
\centering
\includegraphics[width=\textwidth]{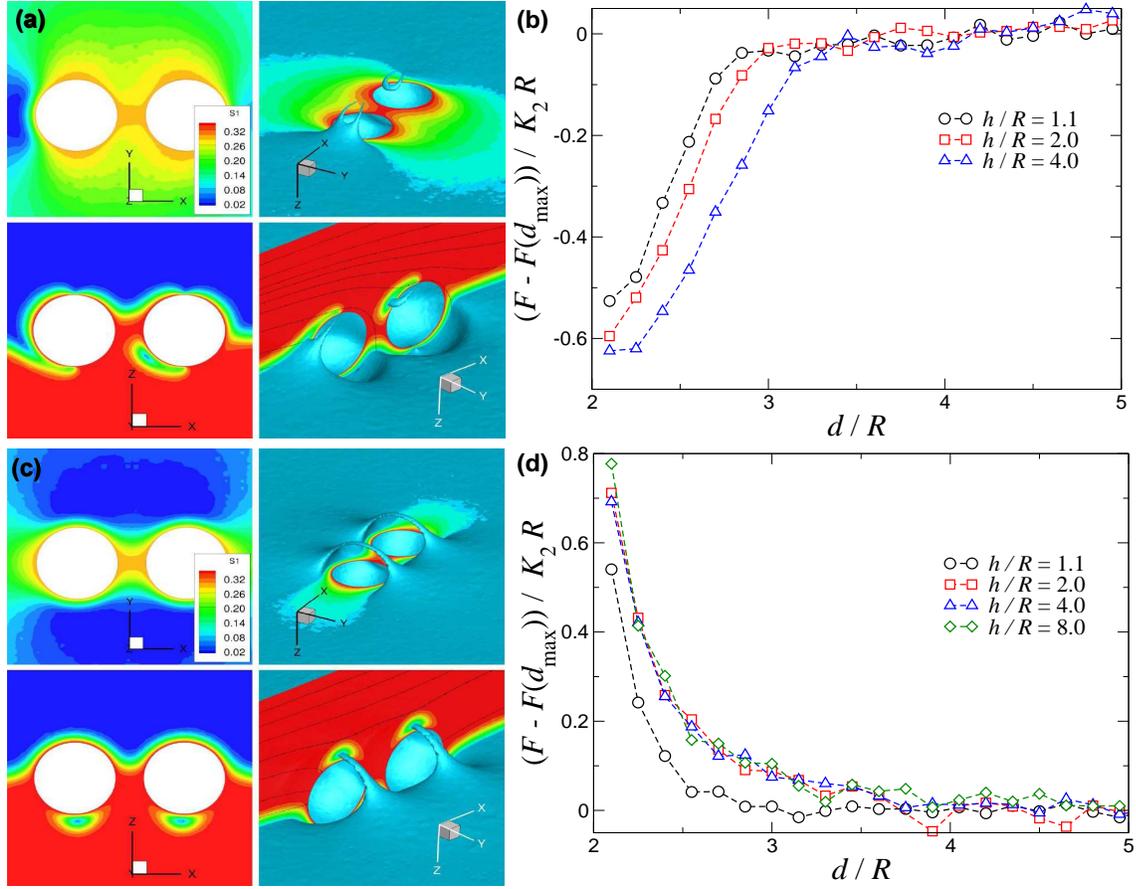}
\caption{Contour plots of the nematic order parameter for (a) planar and (c) homeotropic anchoring of the nematic director on the particle surfaces. The particles are fixed at $z=0$ and separated by distance $d/R=2.7$. The corresponding geometry is depicted in Fig.~\ref{fig:NI_interface_colloids}. Plots (b) and (d) show free energy as a function of distance $d$ between the particles for planar and homeotropic anchoring, respectively. Several film thicknesses are shown. The twist elastic constant, $K_2=9S_{\rm bulk}^2L_1/2$, is evaluated at the bulk NI transition temperature. The values of the material parameters used in Eqs.~(\ref{free_energy_lg}), (\ref{LDG_gradients}) are given in the caption of Fig.~\ref{fig:single}.
\label{fig:double}
}
\end{figure}

The first observation to make is that the structure of the director field, and hence the topological charge of the defects, is similar to the one observed for a single colloidal particle, at least at large separations. The free energy, however, has a very different dependence on the particle-particle separation. In both cases, it is practically insensitive to the film thickness $h$. For homeotropic anchoring, Fig.~\ref{fig:double}d, particles always repel each other. For finite $h$ the director field around a single particle is given asymptotically by the dipole term ~\cite{Oettel2009}, which leads to a dipolar, $(R/d)^3$, decay of the effective interaction at large distances. In our case, the fit of a power law function to the free energy dependencies predicts (qualitatively) a $(R/d)^4$ decay of the interaction potentials, indicating that the situation is more complicated than the theoretical prediction.

A completely different situation is observed for planar anchoring, which is shown in Fig.~\ref{fig:double}b. In this case, particles weakly attract each other with a force which is almost independent of separation. However, below a threshold distance  $d_c\approx 3 R$ which depends on $h$, this weak attraction increases by roughly a factor of ten. In fact, a similar attraction has already been reported earlier and is due to the formation of a nematic bridge or a bridging defect between the particles~\cite{andrienko:2004.b,Andrienko2005}. It effectively leads to an additional force due to the surface tension of the bridge. Capillary bridging leads to a hysteresis in the force-distance profile at small distances~\cite{Andrienko2005} and should not affect the long-range interaction of colloidal particles.

In conclusion, within the Landau-de Gennes formalism we have studied the effective pair interaction between two spherical particles trapped at a NI interface. The interparticle force depends on the type of anchoring as well as the formation,  annihilation, and interaction of defects. Therefore, one  expect that many-body contributions will be important to properly describe collective colloidal ordering at interfaces~\cite{tasinkevych:2006.a,pergamenshchik:2009}.

%% file: conclusions.tex
\section{Summary and outlook}
We have discussed theoretical, computational, and experimental techniques used to study  various inclusions in liquid crystals. In particular, special attention has been paid to two-dimensional systems, such as free-standing smectic films, thin planar nematic cells, and nematic-isotropic interfaces.

From a computational point of view, we find that the Landau-de Gennes free energy functional based on a five component tensorial order parameter is the most suitable description of colloidal systems with anisotropic hosts. By construction, it removes singularities of the director field in the cores of disclination lines or point defects by allowing biaxial perturbations~\cite{LYUKSYUTOV1978} and local melting of the mesophase. It predicts the internal structure of topological defects, is appropriate for describing nematic-isotropic interfaces and suitable for calculating short-range effective interactions between inclusions. The model has  been used extensively to resolve defect structures accompanying a pair of colloidal particles at small separations, to calculate effective interaction potentials and associated binding energies between inclusions, and to study self-assembly of inclusions at liquid-liquid interfaces.

In conclusion, the following issues still remain unresolved: (i) understanding of the temperature dependence and interplay of polarization and elasticity in  SmC* films, leading to rearrangements of surface defects and changes in the type of anchoring at inclusion boundaries; (ii) construction of effective Hamiltonians for interfaces, which can account for bulk degrees of freedom in an effective way to avoid costly calculations using adaptive meshes; (iii) development of adaptive multiscale approaches, where classical density functional or molecular dynamics simulations are used on small scales, for example at the defect core, while a continuum model is used to describe slow variations of an order parameter on larger scales; (iv) development of bottom-up descriptions, which start from a chemical structure and atomistic force-field parametrization and continue with the construction of a coarse-grained particle-based representation~\cite{ruehle:2009.a}, which is then used to parameterize an appropriate free energy functional.